\documentclass[12pt]{article}
\usepackage{epsfig,amsfonts,amsthm}
\newcommand{\be}{\begin{equation}}
\newcommand{\ee}{\end{equation}}
\newcommand{\bea}{\begin{eqnarray}}
\newcommand{\eea}{\end{eqnarray}}
\newcommand{\dst}{\displaystyle}
\newcommand{\fr}[2]{\frac{{\dst #1}}{{\dst #2}}}
\newcommand{\f}{\phi}
\newcommand{\fd}{\phi^\dagger}
\renewcommand{\Re}{\mbox{Re }}
\renewcommand{\Im}{\mbox{Im }}
\newcommand{\stolb}[3]{ \left( \begin{array}{c}#1 \\ #2 \\ #3\end{array}\right) }
\newcommand{\stolbik}[2]{ \left( \begin{array}{c}#1 \\ #2 \end{array}\right) }

\newcommand{\lr}[1]{ \langle #1 \rangle}

\newtheorem{theorem}{Theorem}
\newtheorem{proposition}[theorem]{Proposition}
\newtheorem{lemma}[theorem]{Lemma}

\def\lsim{\mathrel{\rlap{\lower4pt\hbox{\hskip1pt$\sim$}}
    \raise1pt\hbox{$<$}}}         
\def\gsim{\mathrel{\rlap{\lower4pt\hbox{\hskip1pt$\sim$}}
    \raise1pt\hbox{$>$}}}         

\title{Minkowski space structure of the Higgs potential in 2HDM: II.
Minima, symmetries, and topology}

\author{I.P. Ivanov\thanks{E-mail: Igor.Ivanov@ulg.ac.be}\\
  {\small Interactions Fondamentales en Physique et en Astrophysique, Universit\'{e} de Li\`{e}ge,} \\
  {\small All\'{e}e du 6 Ao\^{u}t 17, b\^{a}timent B5a, B-4000 Li\`{e}ge, Belgium}\\
  {\small and}\\
  {\small Sobolev Institute of Mathematics, Koptyug avenue 4, 630090, Novosibirsk, Russia}}

\begin{document}
\maketitle

\begin{abstract}
We continue to explore the consequences of the recently discovered Minkowski space structure
of the Higgs potential in the two-Higgs-doublet model. Here, we focus on the vacuum properties.
The search for extrema of the Higgs potential is reformulated
in terms of 3-quadrics in the 3+1-dimensional Minkowski space.
We prove that 2HDM cannot have more than two local minima in the orbit space
and that a twice-degenerate minimum can arise only via spontaneous violation
of a discrete symmetry of the Higgs potential.
Investigating topology of the 3-quadrics, we give concise criteria
for existence of non-contractible paths in the Higgs orbit space.
We also study explicit symmetries of the Higgs potential/lagrangian
and their spontaneous violation from a wider perspective than usual.
\end{abstract}

\section{Introduction}

\subsection{The Higgs potential in 2HDM and its complexity}

The Standard Model relies on the Higgs mechanism of the
electroweak symmetry breaking. Its simplest realization is based
on a single weak isodoublet of scalar fields, which couple to the
gauge and matter fields and self-interact via the quartic
potential, for review see \cite{Hunter,djouadi1}. Extended
versions of the Higgs mechanisms are based on more elaborate
scalar sectors. The two-Higgs-doublet model \cite{2HDM}, where one
introduces two Higgs doublets $\phi_1$ and $\phi_2$, is one of the
most economic extensions of the Higgs sector beyond the Standard
Model. This model has been extensively studied in literature from
various points of view, see \cite{Hunter,Sanchez,ginzreview,haber,mink}
and references therein. The minimal supersymmetric extension of
the Standard Model (MSSM) uses precisely a specific version of the
2HDM to break the electroweak symmetry, \cite{djouadi2}.

The Higgs potential of the most general 2HDM $V_H = V_2 + V_4$ is
conventionally parametrized as
\bea
V_2&=&-{1\over 2}\left[m_{11}^2(\phi_1^\dagger\phi_1) +
m_{22}^2(\phi_2^\dagger\phi_2)
+ m_{12}^2 (\phi_1^\dagger\phi_2) + m_{12}^{2\ *} (\phi_2^\dagger\phi_1)\right]\,;\nonumber\\
V_4&=&\fr{\lambda_1}{2}(\phi_1^\dagger\phi_1)^2
+\fr{\lambda_2}{2}(\phi_2^\dagger\phi_2)^2
+\lambda_3(\phi_1^\dagger\phi_1) (\phi_2^\dagger\phi_2)
+\lambda_4(\phi_1^\dagger\phi_2) (\phi_2^\dagger\phi_1) \label{potential}\\
&+&\fr{1}{2}\left[\lambda_5(\phi_1^\dagger\phi_2)^2+
\lambda_5^*(\phi_2^\dagger\phi_1)^2\right]
+\left\{\left[\lambda_6(\phi_1^\dagger\phi_1)+\lambda_7
(\phi_2^\dagger\phi_2)\right](\phi_1^\dagger\phi_2) +{\rm
h.c.}\right\}\,.\nonumber
\eea
It contains 14 free parameters: real
$m_{11}^2, m_{22}^2, \lambda_1, \lambda_2, \lambda_3, \lambda_4$
and complex $m_{12}^2, \lambda_5, \lambda_6, \lambda_7$.
Such a large number free parameters makes the analysis of the most general
2HDM and its phenomenological consequences rather complicated.
Even the very first step, finding the minimum of the Higgs potential,
is prohibitively difficult in the most general 2HDM.

On the one hand, in many phenomenological applications one does not
actually need to consider the {\em most} general 2HDM.
Even if one sets several parameters to zero,
there is still room for interesting phenomenology,
and the straightforward algebra is usually sufficient
for the complete treatment of EWSB.

On the other hand, it is obvious that by studying several particular
simplified cases one cannot imagine the {\em full spectrum of possibilities}
offered in 2HDM. This is especially timely now because within few years LHC
is expected to discover experimentally the ESWB mechanism realized in Nature.
In order to safely interpret the LHC data, theorists should know beforehand
which phenomena can or cannot happen in various particular scenarios of EWSB,
in particular, in 2HDM.

Clear view of the general situation in 2HDM will also help understand
which among the free parameters of the Higgs potential are {\em crucial},
in the sense that they shape the phenomenology, and which are {\em redundant},
that is, modify only numerical values of the vacuum expectation values (v.e.v.'s)
of the fields and the Higgs masses.

One particular situation when this knowledge becomes indispensable
is when one attempts to use the existing experimental data to place
bounds on the parameters of 2HDM, for a recent analysis see \cite{per}.
The tricky point here is that there are regions in the parameter space when
the Higgs potential has {\em two different minima} in the orbit space,
see detailed discussion in \cite{twominima,ginzkanishev}.
In this situation one must be aware of the presence of the other minimum, when it exists,
and make sure that the minimum one compares with the data is the global one.
In principle, a method has been developed in which one compares the depth
of the potential at different extrema and related it to some observables such as
mass squared of some Higgs bosons, \cite{twominima,ginzkanishev,depths}.
However, this method gives rather limited information about the structure of the potential;
in particular, it can distinguish minima from saddle points only after substantial
algebraic manipulations.

Finally, experience gained when studying the most general 2HDM
should prove useful, when one turns to even more involved Higgs sectors,
for which the direct algebra with a non-trivial set of parameters becomes
even more difficult.

\subsection{Geometric approaches to the most general 2HDM}

Since the straightforward algebraic calculations are possible
only within very restricted versions of 2HDM, one needs to develop other approaches
how to treat the most general case.
These approaches should aim not at precise analytical calculation
of the v.e.v.'s, Higgs masses etc. (as the algebraic complexity of the most general
2HDM is unavoidable), but at {\em understanding of the general structure in the space of all 2HDM's}.

Since long ago there has been a general understanding that questions of this type
can be answered within a more geometrical rather than analytical approach
to the minimization of a given potential, see \cite{sartori} and references therein.
Back in 1970-1980's, there was much activity on mathematical properties
of various realizations of the symmetry breaking Higgs mechanisms. It was understood that
the problem of minimization of some group-invariant potential is simplified
if one switches from the space of Higgs fields to the {\em orbit space}, \cite{michel}.
This idea was exploited in \cite{sartori,Kim} to study the minima of
a Higgs potential invariant under the Lie group $G$ with Higgs fields transforming
under various representations of this group. This general approach has been
even applied to 2HDM, see \cite{sartori2hdm}.

Last several years witnessed a renewed interest in the study of the most general 2HDM.
The idea was to exploit the {\em reparametrization properties} of the 2HDM potential,
rewriting (\ref{potential}) as convolution of some second and fourth-rank Higgs field tensors
with the corresponding tensors constructed from the free parameters available, \cite{CP}.
However, the machinery based on this tensorial approach, \cite{haber}, lacked transparency
and intuition, which was nicely illustrated by its application to the
problem of $CP$-violation in 2HDM, \cite{haber2}.

These drawbacks were avoided in the group-theoretic/linear algebraic approach of \cite{group}.
This approach put together the benefits of the tensorial and geometric formalisms
by discovering some simple structure in the tensors used in the former.
This idea was developed further in \cite{mink} by considering the largest
reparametrization group of the Higgs potential, $GL(2,C)$, and observing
that its subgroup $SL(2,C)$ induces the Minkowski space structure in the orbit space of 2HDM.
Ref.~\cite{mink} showed the prominent role played by
the lightcone and some caustic surfaces in this Minkowski space.

The linear algebraic/geometric properties of the 2HDM were also studied in \cite{nishi}
and were later extended to the general $N$-Higgs-doublet models in \cite{nishiN}.
The geometric point of view was also used in \cite{nachtmann2} to study the $CP$-violation
in 2HDM.

\subsection{Plan of the paper}

The plan of the paper is the following. In Section~\ref{section-mink} we
review the Minkowski-space formalism introduced in \cite{mink}.
In Section~\ref{sectequipotential} we reformulate the minimization problem
of the Higgs potential in terms of geometry of 3-quadrics embedded in the Minkowski space.
We prove there that the 2HDM, if it has a discrete set of minima, can have no more than two
local minima.
We also introduce there the valley of the Higgs potential and
discuss the consequences of its non-trivial topology.
Section~\ref{section-symmetries} deals with discrete symmetries of the Higgs
lagrangian/potential as well as their spontaneous violation
from a somewhat more general point of view than usual.
Then, in Section~\ref{section-global}, we take a closer at the situation with
a doubly degenerate global minimum. We draw conclusions in Section~\ref{section-conclusions},
and in Appendix we give some useful formulas and
prove the lemma that we use in Propositions~\ref{prop-degenerate-vacuum}
and~\ref{prop-degenerate-vacuum2}.

We find it useful to summarize here, in plain words, the main results of this paper:
\begin{itemize}
\item
The search for the global minimum of the Higgs potential is equivalent
to the search of a second order 3D surface that touches but never intersects
the future lightcone $LC^+$ in the Minkowski space.
\item
The geometry of this and similar surfaces, which is related in a very transparent way
to the parameters of the potential, plays an important role
in various phenomena in the scalar sector of 2HDM. For example,
we prove that the (tree-level) Higgs potential cannot have more than two local minima
and that a doubly degenerate vacuum can appear only as a result
of spontaneous violation of a specific reparametrization symmetry
of the potential. These surfaces can also have non-trivial topology
and give rise to non-contractible paths in the Higgs orbit space,
leading possibly to metastable quasi-topological excitations within the scalar sector
of 2HDM.
\item
We list all reparametrization symmetries the Higgs lagrangian can
have, underline difference between symmetries of the potential and
of the whole Higgs lagrangian, and establish that the maximal
spontaneous violation of a discrete symmetry in 2HDM consists in
removing one $Z_2$ factor.
\end{itemize}

\section{Minkowski space structure of the orbit space of the 2HDM}\label{section-mink}

Here we briefly review the Minkowski space formalism introduced in \cite{mink}
and remind some of the results obtained there.

\subsection{Extended reparametrization group}\label{subsection-rep}

The starting point is the organization of the Higgs doublets $\phi_1$ and $\phi_2$
into a hyperspinor $\Phi$:
$$
\Phi = \stolbik{\phi_1}{\phi_2}\,.
$$
The key observation then is that the potential (\ref{potential}) retains its generic form
under any linear transformation between doublets $\phi_1$ and $\phi_2$.
In other words, (\ref{potential}) is {\em invariant} under a linear transformation
of $\Phi$ accompanied with an appropriate transformation of the parameters $\lambda_i$
and $m_{ij}^2$ (see detailed discussion in \cite{mink}).

Let us introduce the four-vector $r^\mu = (r_0,\,r_i) = (\Phi^\dagger \sigma^\mu \Phi)$ with components
\be
r_0 = (\Phi^\dagger \Phi) = (\fd_1 \f_1) + (\fd_2 \f_2)\,,\quad
r_i = (\Phi^\dagger \sigma_i \Phi) =
\stolb{(\fd_2 \f_1) + (\fd_1 \f_2)}{-i[(\fd_1 \f_2) - (\fd_2 \f_1)]}{(\fd_1 \f_1) - (\fd_2 \f_2)}\,.
\label{ri}
\ee
The quantities $r_0$ and $r_i$ are gauge-invariant as they do not
change when the electroweak gauge transformations
act on $\phi_1$ and $\phi_2$ simultaneously. Thus, $r^\mu$ parametrizes {\em gauge orbits}
of the Higgs fields.
The $SL(2,C)$ group of transformations of the spinor $\Phi$ induces
the proper Lorentz group $SO(1,3)$ of transformations of $r^\mu$.
Thus, the orbit space in which the Higgs potential is defined
is equipped with the Minkowski space structure.

An important remark is in order.
Since the Higgs fields are operators, the quantity $r^\mu$ is an operator-valued
four-vector. However, in the present investigation we are not interested in the dynamics
of 2HDM  but aim only at the understanding of the vacuum structure of 2HDM. To this end,
we will be interested not with $r^\mu$ itself, but with its vacuum expectation value
$\lr{r^\mu}$, which is a $c$-number.

We call this $SO(1,3)$ transformation group the extended reparametrizaton group.

Now, since $\lr{r_0} \ge 0$ and, due to the Schwartz lemma, $\lr{r^\mu} \lr{r_\mu} \equiv \lr{r_0}^2 - \lr{r_i}^2 \ge 0$,
the space of all possible orbits, the {\em orbit space}, is given by the
forward lightcone $LC^+$ in the Minkowski space. The extended reparametrization group
in the orbit space, $SO(1,3)$, leaves the orbit space invariant.
Note that $\lr{r^\mu} \lr{r_\mu}$ coincides with the quantity $Z$ introduced
in \cite{ginzreview,ginzkanishev}.

The Higgs potential in the orbit space can be written in a very compact form:
\be
V = - M_\mu r^\mu + {1\over 2}\Lambda_{\mu\nu} r^\mu r^\nu\,,\label{Vmunu}
\ee
where
\bea
M^\mu &=& {1\over 4}\left(m_{11}^2+m_{22}^2,\, -2\Re m_{12}^2,\,
2\Im m_{12}^2,\, -m_{11}^2+m_{22}^2\right)\,,\nonumber\\[2mm]
\Lambda^{\mu\nu} &=& {1\over 2}\left(\begin{array}{cccc}
{\lambda_1+\lambda_2 \over 2} + \lambda_3 & -\Re(\lambda_6 + \lambda_7)
    & \Im(\lambda_6 + \lambda_7) & -{\lambda_1-\lambda_2 \over 2} \\[1mm]
-\Re(\lambda_6 + \lambda_7) & \lambda_4 + \Re\lambda_5 & -\Im\lambda_5 & \Re(\lambda_6 - \lambda_7) \\[1mm]
\Im(\lambda_6 + \lambda_7) & -\Im\lambda_5 & \lambda_4-\Re\lambda_5 & -\Im(\lambda_6 - \lambda_7) \\[1mm]
 -{\lambda_1-\lambda_2 \over 2} & \Re(\lambda_6 - \lambda_7) & -\Im(\lambda_6 - \lambda_7)
 & {\lambda_1+\lambda_2 \over 2} - \lambda_3
\end{array}\right)\,.\label{lambda}
\eea
We repeat again that when searching for the minima of the potential, we will understand $r^\mu$ in (\ref{Vmunu}) in the sense
of vacuum expectation values.

Properties of $\Lambda_{\mu\nu}$ were explored in \cite{mink}. It was shown that
if one requires the Higgs potential to be positive-definite at
large quasiclassical values of the Higgs fields, then $\Lambda_{\mu\nu}$
is positive-definite on and in the forward lightcone $LC^+$.
This is equivalent to the statement that $\Lambda_{\mu\nu}$ is diagonalizable by an $SO(1,3)$ transformation
and after diagonalization it takes form
$$
\left(\begin{array}{cccc}
\Lambda_0 & 0 & 0 & 0\\
0 & -\Lambda_1 & 0 & 0\\
0 & 0 & -\Lambda_2 & 0\\
0 & 0 & 0 & -\Lambda_3 \end{array}\right)\quad \mbox{with}\quad
\Lambda_0 > 0\ \mbox{and} \ \Lambda_0 > \Lambda_1, \Lambda_2, \Lambda_3\,.
$$
We will refer to $\Lambda_0$ as the ``timelike'' eigenvalue of $\Lambda_{\mu\nu}$
and $\Lambda_i$, $i=1,2,3$, as its ``spacelike'' eigenvalues.
For the reader's convenience, we collect in Appendix~\ref{appendixlambda} some basic
formulae concerning manipulations with $\Lambda_{\mu\nu}$.

In principle, one can slightly relax the above condition by requiring $\Lambda_{\mu\nu}$
to be {\em non-negative} instead of positive-definite within $LC^+$.
This implies the possibilities of $\Lambda_0 = 0$
and/or $\Lambda_i = \Lambda_0$ for some $\Lambda_i$. These possibilities lead to existence
of ``flat'' directions of $\Lambda_{\mu\nu}$ and are viable only when the mass term $-M_\mu r^\mu$ grows along
these directions.

\subsection{Extrema of the Higgs potential}

Minimization (or in general, extremization) of the potential in the Higgs space
amounts to finding the minimum (extremum) of (\ref{Vmunu})
on or inside the future lightcone $LC^+$. It can be easily shown that 2HDM potential
bounded from below cannot have nontrivial maxima, so all nontrivial extrema
are either minima or saddle points \cite{michel,mink}.

If the minimum lies on the surface of $LC^+$,
the v.e.v.'s of the Higgs doublets can be brought by an appropriate gauge transformation
to the standard form
\be
\langle \phi_1\rangle  = {1\over\sqrt{2}}\stolbik{0}{v_1},\quad
\langle \phi_2\rangle  = {1\over\sqrt{2}}\stolbik{0}{v_2 e^{i\xi}} \,,\label{vev1}
\ee
with real $v_1$, $v_2$, $\xi$. This corresponds to the {\em neutral vacuum}, since
it remains invariant under residual $U(1)_{EM}$ gauge transformations
and the photon remains massless.
If the minimum lies strictly inside $LC^+$, then gauge transformations can bring the v.e.v.'s to
\be
\langle \phi_1\rangle  = {1\over\sqrt{2}}\stolbik{0}{v_1},\quad
\langle \phi_2\rangle  = {1\over\sqrt{2}}\stolbik{u}{v_2 e^{i\xi}}\label{vev2}
\ee
with some nonzero real $u$. This situation corresponds to the {\em charge-breaking vacuum}
with massive photon.

The condition for the extremum strictly inside $LC^+$ is
\be
\Lambda^{\mu\nu} \lr{r_\nu} = M^\mu\,.\label{extremum1}
\ee
For non-singular $\Lambda_{\mu\nu}$, it always exists and is unique. However,
it is realizable as a Higgs field configuration only if
$m_\mu = (\Lambda^{-1})_{\mu\nu}M^\nu$ lies inside $LC^+$.

The condition for the extrema lying on the surface of $LC^+$ are written with the aid of
a Lagrangian multiplier $\zeta$:
\be
\Lambda^{\mu\nu} \lr{r_\nu} - \zeta \cdot \lr{r^\mu} = M^\mu\,.\label{extremum2}
\ee
In general, there can be up to six neutral extrema.
In ref.~\cite{mink} it was shown that the sign of $\zeta$ determines
the sign of the mass square of the charged degrees of freedom.
Thus, one of the necessary condition for a neutral extremum to be
minimum is $\zeta > 0$. Geometrically, it means that the potential increases
as one shifts from the surface of $LC^+$ inwards, or in other words, that the mass square of
the charged excitations is positive.

In \cite{mink} we found a simple criterion when the Higgs potential has charge-breaking global minimum
and proved the theorem that neutral, and charge-breaking minima cannot coexist
in 2HDM.

\subsection{Non-standard kinetic term}

Transformations from the extended reparametrization group modify the Higgs kinetic term.
However, it can also be rewritten in the explicitly reparametization-covariant form:
\be
K = \rho^\mu K_\mu\,,\quad \rho^\mu = (D_\alpha \Phi)^\dagger \sigma^\mu (D^\alpha \Phi)\,,
\ee
where $D_\alpha$ is the extended derivative, $\alpha$ denotes the usual space-time coordinates,
while $\mu$, as before, denotes the coordinate in the Higgs orbit space.
Note that reparametrization transformation properties of $\rho^\mu$ are the same as $r^\mu$.
The entire Higgs lagrangian is simply $L = K - V$. In the
usual frame, the ``kinetic'' four-vector $K_\mu$ is simply $K_\mu = (1,\,0,\,0,\,0)$.
Boosts make $K_\mu$ a non-trivial vector, but it always obeys $K^\mu K_\mu = 1$ and
always lies inside the future lightcone.

Having non-standard kinetic term represents only a minor inconvenience when one studies
the {\em general structure} of the Higgs potential. The number of extrema, their minimum/saddle point
classification, the depth of the potential are all insensitive to the non-standard kinetic term.
It is only the exact numerical value of the v.e.v. and masses of the physical Higgs bosons
that do depend on $K_\mu$. Non-standard kinetic term also leads to distinction between
the symmetries of the potential and of the entire Higgs lagrangian,
which will be discussed in Section~\ref{section-symm-pot-lang}.

\subsection{Prototypical model and the degree of algebraic complexity of 2HDM}
\label{section-prototypical}

The extended reparametrization symmetry of the Higgs potential reduces the number of
the crucial parameters of the potential.

The diagonalizability of $\Lambda_{\mu\nu}$ means that for any generic 2HDM
upon performing a suitable linear transformation of the Higgs doublets one can arrive at
the Higgs potential with parameters $\bar\lambda_i$, which satisfy the following relations:
\be
\bar\lambda_1=\bar\lambda_2\,,\quad \bar\lambda_6=\bar\lambda_7=0\,,\quad \Im \bar\lambda_5 = 0\,,
\ee
together with a generic set of $\bar m^2_{ij}$ and a generic kinetic term.
We call it the {\em prototypical model} of a given 2HDM.
The structure of the extrema (the number of the extrema,
their minimum/saddle point classification, their depth and symmetries)
of the original Higgs potential are the same as for the prototypical model and
depends only on 7 parameters: the four eigenvalues of $\Lambda_{\mu\nu}$ and
the three ratios of the components of $M_\mu$ that define its direction in the Minkowski space
in the prototypical model.

In the geometric treatment of the Higgs potential in a generic 2HDM we manipulate with
the eigenvalues of $\Lambda_{\mu\nu}$ and components $M_\mu$ of the prototypical model.
If one intends to obtain these values from the initial generic set of $\lambda_i$ and
$m_{ij}^2$, one has to solve the fourth-order characteristic equation.
One can say that {\em the degree of algebraic complexity} of a generic 2HDM is four.

In special cases, when $\Lambda_{\mu\nu}$ is already block-diagonal, this degree is lower.
For example, in the often-considered case $\lambda_6=\lambda_7=0$, $\Lambda_{\mu\nu}$
is made of two blocks $2\times 2$. Its degree of complexity is 2, and in order to diagonalize
$\Lambda_{\mu\nu}$ one has to perform, independently, a boost along third axis a rotation in the
``transverse'' plane. This makes such a model tractable with the straightforward calculations.

For the sake of illustration, let us note that in the tree-level MSSM $\Lambda_{\mu\nu}$
is already diagonal, with the following eigenvalues:
\be
\Lambda_0 = 0\,\quad \Lambda_1 = \Lambda_2 = - g_2^2 \,\quad \Lambda_3 = - {g_1^2+g_2^2 \over 2}\,,
\ee
where $g_1$, $g_2$ are the EW gauge coupling constants.
As discussed at the end of Section~\ref{subsection-rep}, $\Lambda_0 = 0$ is possible
but it requires that $M_0<0$, which is indeed satisfied in the tree-level MSSM.

\section{Equipotential surfaces, minima, and the valley of the Higgs potential}\label{sectequipotential}

Let us continue our investigation of the consequences of the Minkowski space structure
of the orbit space of the 2HDM.

First, we introduce some notation.
Let ${\cal M}$ be Minkowski space of all possible four-vectors $p^\mu$. As it was noted above,
only vectors lying on and inside the future lightcone $LC^+$
are physically realizable via Higgs fields (\ref{ri}).
Choose a vector $p^\mu$ from ${\cal M}$ and consider quadratic form $\Lambda_{\mu\nu} p^\mu p^\nu$.
Upon diagonalization of $\Lambda_{\mu\nu} p^\mu p^\nu$ by an appropriate $SO(1,3)$ transformation,
one can rewrite the quadratic form as
\be
\Lambda_{\mu\nu} p^\mu p^\nu = \Lambda_0 p_0^2 - \sum_i \Lambda_i p_i^2\,. \label{Lampp}
\ee
Due to the properties of $\Lambda_{\mu\nu}$, this quadratic form is positive definite if $p^\mu$ lies in the future lightcone,
but it is not required to be positive definite in the {\em entire} Minkowski space ${\cal M}$.

Let us define the 3-manifold ${\cal M}^0$ as the locus of all $p^\mu$ such that $\Lambda_{\mu\nu} p^\mu p^\nu = 0$.
In addition, we also denote by ${\cal M}^+$ and ${\cal M}^-$ the parts of the entire Minkowski space, where this quadratic form
is positive and negative, respectively. Clearly, ${\cal M}^0$ separates ${\cal M}^+$ and ${\cal M}^-$.

More generally, we introduce a 3-manifold ${\cal M}^C$ as the locus of all $p^\mu$ such that $\Lambda_{\mu\nu} p^\mu p^\nu = C$,
which separates ${\cal M}$ into regions ${\cal M}^{<C}$ and ${\cal M}^{>C}$.
Note that 3-manifolds ${\cal M}^C$ are {\em nested}, in the sense that they never intersect
and ${\cal M}^{C_1}$ lies in ${\cal M}^{>C_2}$ if $C_1>C_2$.

\subsection{Geometry of 3-manifolds ${\cal M}^C$}\label{sectiongeometryMC}
Let us now study the geometry of a typical 3-manifold ${\cal M}^C$.
As can be seen from (\ref{Lampp}), it is a second-order 3-surface (3-quadric) embedded
in the 4D space. More specifically,
it is a 3-hyperboloid (or a 3-cone for $C=0$), whose shape depends on the sign of $C$ and of $\Lambda_i$.
Let us list explicitly all the cases.

\begin{figure}[!htb]
   \centering
\includegraphics[width=6.5cm]{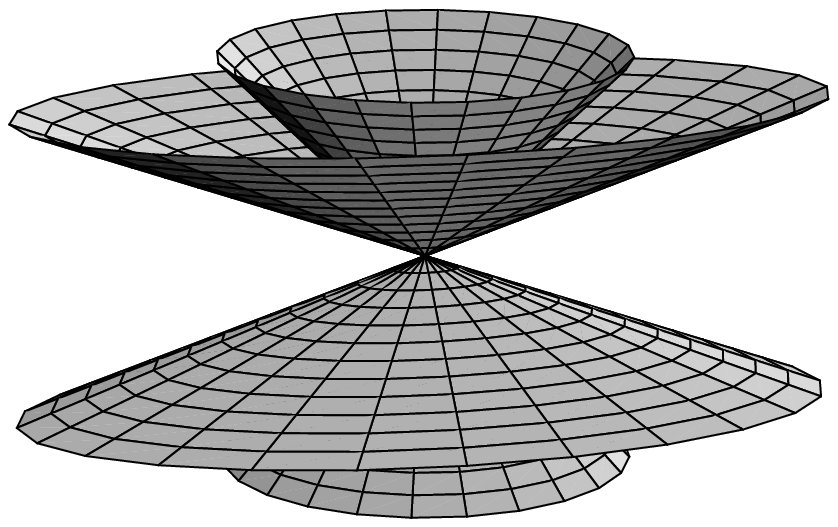}\hspace{2cm}
\includegraphics[width=5cm]{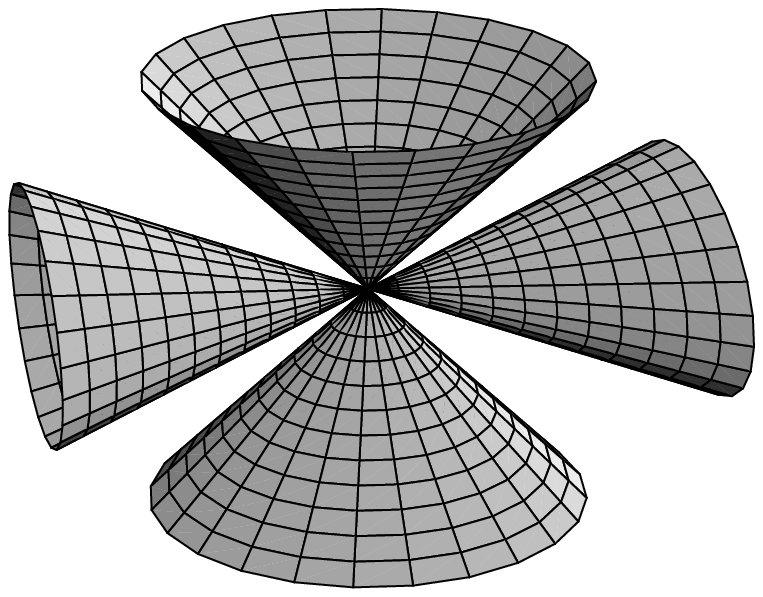}
\caption{Examples of the manifolds ${\cal M}^0$ in the $2+1$-dimensional picture. Left pane: all $\Lambda_i>0$,
right pane: $\Lambda_1>0$, $\Lambda_2<0$. In each case, the lightcone $LC^+$ is also shown for comparison.}
   \label{fig-M0}
\end{figure}

\begin{itemize}
\item {\bf All $\Lambda_i$ are positive}.
3-quadric ${\cal M}^0$ is a pair of 3D cones oriented along the ``time-like" axis.
Note that due to $\Lambda_0>\Lambda_i$ these cones are wider than the lightcone $LC$.
The interior of these cones is ${\cal M}^+$. To help the reader visualize this construction,
we show in Fig.~\ref{fig-M0}, left, the $2+1$-dimensional analogs of ${\cal M}^0$ together with $LC$.
A 3-quadric ${\cal M}^C$ with $C>0$ or $C<0$ is a two-sheet or one-sheet 3-hyperboloid, respectively,
oriented along the ``time-like" axis.

\item {\bf $\Lambda_1,\,\Lambda_2>0$, while $\Lambda_3<0$}.
${\cal M}^0$ is a peculiar cone, specific for a higher dimensional space, defined by equation:
$$
\Lambda_1 p_1^2 + \Lambda_2 p_2^2 - |\Lambda_3| p_3^2 - \Lambda_0 p_0^2 = 0\,.
$$
${\cal M}^C$ are similarly peculiar one-sheet 3-hyperboloids.

\item {\bf $\Lambda_1>0$ while $\Lambda_2,\,\Lambda_3<0$}.
${\cal M}^0$ is now a pair of cones, similar to the all-positive case, but oriented along the first spacelike,
rather than timelike, axis. Again, we illustrated this case in Fig.~\ref{fig-M0}, right,
with the $2+1$-dimensional analogs.
Its interior now is ${\cal M}^-$.
${\cal M}^C$ with negative $C$ lie inside this cone and are two-sheet 3-hyperboloids, again oriented along
the first axis. ${\cal M}^C$ with positive $C$ are one-sheet 3-hyperboloids.

\item {\bf All $\Lambda_i$ are negative}.
In this case $\Lambda_{\mu\nu}$ is positive definite in the entire Minkowski space, so ${\cal M}^0$
is reduced to the single point at origin, $p^\mu = 0$.
3-surfaces ${\cal M}^C$ with negative $C$ do not exist, while
${\cal M}^C$ with positive $C$ are 3-ellipsoids defined by
$$
|\Lambda_1| p_1^2 + |\Lambda_2| p_2^2 + |\Lambda_3| p_3^2 + \Lambda_0 p_0^2 = C\,.
$$

\item
{\bf If there is a zero among $\Lambda_i$}, e.g. $\Lambda_k = 0$,
then the above 3-manifolds ${\cal M}^C$ become cylindric along the $k$-th axis.
\end{itemize}

\subsection{Relation to the minimization problem}

Let us now demonstrate the following simple geometric fact:\\

\noindent {\em the search for the neutral extrema of the Higgs potential can be always reformulated
as the search for such 3-quadrics that touch the forward lightcone $LC^+$.}\\

Let us first assume that $\Lambda_{\mu\nu}$ is non-singular, i.e. its eigenvalues $\Lambda_i \not = 0$.
Then $\Lambda^{-1}_{\mu\nu}$ exists, and one can rewrite the Higgs potential (\ref{Vmunu}) as
\be
V = {1\over 2}\Lambda_{\mu\nu} (r^\mu-m^\mu) (r^\nu-m^\nu) + V_0\,,\quad
m_\mu = (\Lambda^{-1})_{\mu\nu}M^\nu\,,\quad
V_0 = -{1\over 2}(\Lambda^{-1})_{\mu\nu}M^\mu M^\nu\,.
\label{Vmunu2}
\ee
Let us now denote $p^\mu=r^\mu-m^\mu$. Then, the 3-surface ${\cal M}^C$ is in fact the surface
of equal values of the potential, $V=V_0+C/2$. Intersection of ${\cal M}^C$ with (the surface and interior of)
the future lightcone $LC^+$ defines the corresponding {\em equipotential 3-surface}.

Note that the 3-manifolds ${\cal M}^C$ are constructed starting from the base point $r^\mu = m^\mu$.
Therefore, ${\cal M}^C$ are shifted from $LC$, and the shape of their intersection can be non-trivial.

\begin{figure}[!htb]
   \centering
\includegraphics[width=7cm]{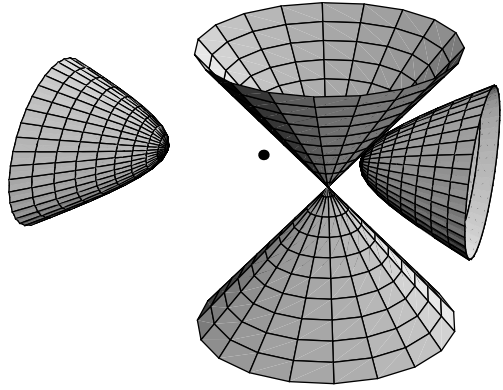}
\caption{A $2+1$-dimensional illustration of the contact of ${\cal M}^{C_\mathrm{min}}$ with $LC^+$.
The thick point indicates the position of $m^\mu$.}
   \label{fig-contact}
\end{figure}

Minimization of the Higgs potential, therefore, amounts to finding the minimal value of $C$, $C_\mathrm{min}$, for
which the equipotential surface exists. This 3-surface, which we label as ${\cal M}^{C_\mathrm{min}}$,
{\em never intersects but only touches}\footnote{Here by ``touch'' we mean that
the two 3-manifolds not only pass through this point, but also have
parallel normals at this point.} $LC^+$.
This is what makes ${\cal M}^{C_\mathrm{min}}$ unique among all ${\cal M}^C$.
To facilitate the visualization, Fig.~\ref{fig-contact} shows a $2+1$-dimensional example
of the contact between ${\cal M}^{C_\mathrm{min}}$ and $LC$ for a specific case when $\Lambda_1>0$,
$\Lambda_2 <0$ and $m^\mu$ lying outside $LC^+$.

Since 3-surfaces ${\cal M}^C$ are nested,
all ${\cal M}^C$ with $C<C_\mathrm{min}$ form a region ${\cal M}^{<C_\mathrm{min}}$ in ${\cal M}$,
which is disjoint from $LC^+$, while all 3-surfaces ${\cal M}^C$ with $C>C_\mathrm{min}$,
forming region ${\cal M}^{>C_\mathrm{min}}$, intersect $LC^+$.
Among the latter there might be other 3-quadrics that in addition to intersection also touch $LC^+$
at some point. These are the other extremal 3-manifolds, which correspond either to the
local minimum or a saddle point.

Let us now consider the case of singular $\Lambda_{\mu\nu}$. To consider a concrete example, suppose that
only $\Lambda_3=0$. If $M_3 \not = 0$, then the above shift of the base point cannot be used in its initial form.
Instead consider this shift in the subspace where $\Lambda_{\mu\nu}$ is not singular:
\be
V = {1\over 2}\left[\Lambda_0(r_0-m_0)^2 - \Lambda_1(r_1-m_1)^2 - \Lambda_2(r_2-m_2)^2)\right] + M_3 r_3 + \bar{V}_0\,,
\label{Vmunu3}
\ee
Note that $V$ is now linear, not quadratic, in $r_3$.
Thus, a generic ${\cal M}^C$ (whose definition now includes the $M_3 r3$ term) is now a 3-paraboloid
with one spacelike parabolic direction. So, here again, the search for the stationary points of the potential
is cast into the form of finding paraboloids that touch the forward lightcone $LC^+$.
The case of $\Lambda_0 = 0$, $\Lambda_i <0$ is analyzed in the similar way. The generic ${\cal M}^C$
is again an elliptical 3-paraboloid with the timelike parabolic direction.\\

\subsection{The number of local minima}

Let us now apply the above constructions to the question of the number of local minima, at the tree-level,
in the most general 2HDM. We will first consider one very particular case,
show that there can be no more than two local minima, and then prove that
this number bounds also the generic situation.\\

We start with a special case of 2HDM with $\Lambda_{\mu\nu}$, whose eigenvalues $\Lambda_i$
are all positive and distinct, and with
$M_\mu$ lying on the future direction (in the diagonal basis): $M_\mu = (M_0,\,0,\,0,\,0)$.
This situation can be treated with the straightfroward algebra
(in the diagonal basis), but it is instructive to study this case
geometrically.

Consider first the ``horizontal'' 3-section at some positive $r_0$
of the construction described in the previous subsection (i.e. the Lightcone $LC^+$ and the family
of 3-manifolds ${\cal M}^C$ constructed at the base point $m^\mu$).
Then, rescale all the spacelike coordinates by introducing $\tilde{r}_i = r_i/r_0$.
Then, in the $\tilde{r}_i$ space, the 3-section of the surface of $LC^+$
is always the unit sphere, while the 3-sections of ${\cal M}^C$ are ellipsoids,
with the same symmetry center as the sphere.
If ${\cal M}^C$ is an extremal 3-surface, then this ellipsoid
touches sphere in two opposite points.

Now, consider any of the three two-dimensional sections inside this 3-section
that passes through the common symmetry point and is parallel to two $\Lambda_{\mu\nu}$'s
eigenvectors, say, $e_1$ and $e_2$. The 2-section of $LC^+$ is then the unit circle,
while the section of ${\cal M}^C$ is an ellipse.

\begin{figure}[!htb]
   \centering
\includegraphics[width=17cm]{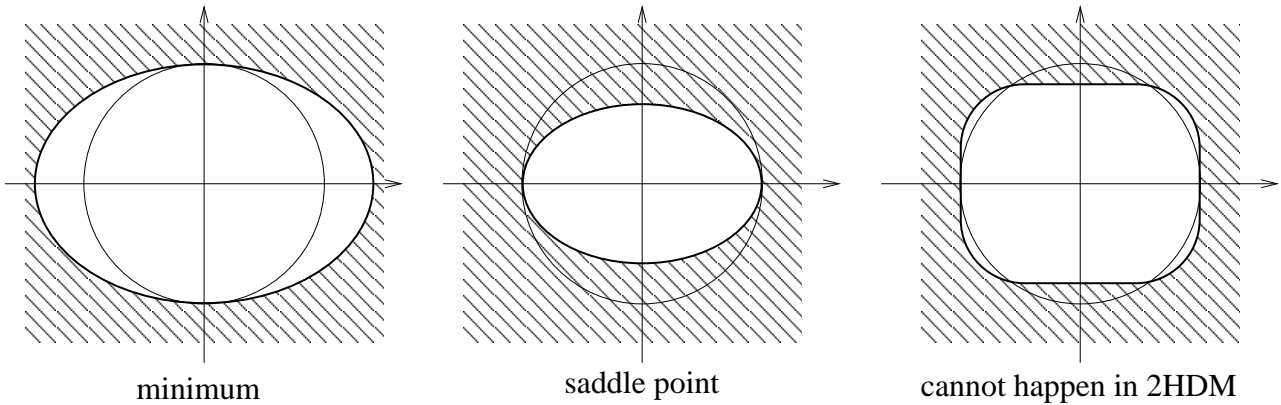}
\caption{The possible 2-sections of ${\cal M}^C$ and $LC^+$ for the case when all $\Lambda_i>0$.
The left and middle plots correspond to the minimum and the saddle point, respectively, while
the right plot cannot happen in 2HDM.}
   \label{fig-ellipse}
\end{figure}

If ${\cal M}^C$ is an extremal 3-surface and if the contact points belong to this 2-section,
then one can have either of the first two situations depicted in Fig.~\ref{fig-ellipse}.
Here, the circle is the 2-section of $LC^+$, the ellipse shown in the thick line
is the section of ${\cal M}^C$ and the shaded area is the section of ${\cal M}^{<C}$.

The fact that the section of ${\cal M}^C$ is an ellipse, i.e. a {\em second-order curve},
makes the intersection shown in Fig.~\ref{fig-ellipse}, right, impossible in 2HDM.
Indeed, two second-order curves
can have at most four intersection points or at most two contact points (here, by ``contact'' or
``touch'' we mean again a 2-point intersection). This means that within the symmetric geometry we consider,
if the circle touches the ellipse in two points, then it must lie completely inside
(Fig.~\ref{fig-ellipse}, left) or completely outside (Fig.~\ref{fig-ellipse}, middle) the ellipse.

In order to understand whether a given configuration corresponds to the minimum or a saddle point,
one must shift away from the contact point in all directions still staying on the surface of $LC^+$.
If one gets into the shaded region, then this shift happens to minimize the potential further,
so it cannot be the minimum. This is so for Fig.~\ref{fig-ellipse}, middle, but not
for Fig.~\ref{fig-ellipse}, left.

Repeating this analysis for all the 2-sections, one arrives at the conclusion that
there exists only one ellipse that corresponds to the minima, the one that lies completely inside
the ellipsoid. The other correspond to the saddle-points.
It means that there are at most two minima in the orbit space in this special version of 2HDM.

Now, the same inspection can be repeated for 2HDM with non-positive $\Lambda_i$. In this case
one will encounter not only ellipses but also hyperbolas and parabolas (if some $\Lambda_i$ are zeros).
In all these cases one finds that the above conclusion --- {\em there are at most two minima} ---
still holds.

The only exception is when some of $\Lambda_i > 0$ coincide. In this case, the 2-section
of ${\cal M}^C$ generated by the corresponding eigenvectors will be not the ellipse but the circle,
and can lead to a {\em continuum of minima}. So, the above conclusion of two minima applies to the
situations when the number of minima is finite.\\

The second step is to prove that the above analysis with the very special choice of $M_\mu$
is in fact representative of {\em the most general situation} with largest possible number of extrema.

To this end, we recall some relevant results from \cite{mink}.
The space of all possible four-vectors $M_\mu$ is naturally broken into regions
with different number of extrema. The 3-separatrices of these regions are
the forward $LC^+$ and backward $LC^-$ lightcones as well as up to two caustic surfaces.
These caustic surfaces are aligned with the timelike eigenvector of $\Lambda_{\mu\nu}$.
The number of neutral extrema (in the orbit space)
was given by Proposition~7 of \cite{mink}:
\begin{enumerate}
\item if $M_\mu$ lies outside $LC^-$, at least one neutral extremum exists;
\item if $M_\mu$ lies inside $LC^+$, at least two neutral extrema exist;
\item if $M_\mu$, in addition, lies inside one of the caustic cones, two neutral extrema appear, in addition to criteria (1) or (2);
\item if $M_\mu$ lies inside both caustic cones, four neutral extrema appear, in addition to criteria (1) or (2).
\end{enumerate}
The key feature is that the spectrum (the number and the minimum/saddle point nature) of stationary points
remains the same for all $M_\mu$ inside any given region.
In order to change the number or nature of the stationary points,
one must cross the 3-separatrix.

In particular, the entire innermost region (with $M_\mu$ lying inside $LC^+$
and both caustic cones) has the same spectrum of extrema no matter what representative $M_\mu$ one chooses.
Let us choose $M_\mu$ along the future direction: $M_\mu = (M_0,\,0,\,0,\,0)$.
Then the spectrum of extrema in this case (with generic non-equal values of $\Lambda_i$)
will represent the largest possible number of extrema in the orbit space:
6 neutral plus one charge-breaking saddle point plus one EW symmetric maximum at the origin.
Our analysis tells that no more than two of them are minima.
This completes the prove of the following statement:
\begin{proposition}\label{prop-twominima}
The most general 2HDM with a discrete set of minima can have at most two local minima.
\end{proposition}

Note that the number of local minima in 2HDM (at the tree-level) was discussed recently in \cite{twominima}.
There, authors use the straightforward algebra together with the Morse theory and analyze the number
of stationary points and, in particular, minima of the 2HDM. Unfortunately, they work not in the orbit space
but deal with the typical representatives of these orbits, which sometimes leads to double counting.

In particular, they argued that two pairs of degenerate minima plus a minimum at origin can take place in 2HDM
upon a suitable choice of parameters.
Even if each of these pairs corresponds to a single orbit, this statement would imply existence of three
minima in 2HDM, which contradicts the Proposition we just proved\footnote{In fact, it can be shown by simple arguments
that the minimum at origin cannot coexist with any other stationary point.}.

In addition, authors of \cite{twominima} found, by extensive numerical search,
that it is possible to have coexisting $CP$-conserving and spontaneously $CP$-violating minima in 2HDM,
although, as they say, ``the combination of parameters coresponding to this situation are extremely rare''.
Since spontaneous $CP$-violating minima always come in pairs, this also implies existence of three distinct
minima in the orbit space, which again contradicts the above Proposition.
In fact, even a more general statement follows from Proposition~\ref{prop-twominima}:\\

\noindent {\bf Corollary:}
{\em Whatever the discrete symmetry of the Higgs potential is, minima that conserve and violate
this symmetry cannot coexist in 2HDM.}\\

The fact that $CP$-conserving and spontaneous $CP$-violating minima cannot coexist in 2HDM was noted also
in \cite{ginzkanishev}.
The more general statement proved in Proposition~\ref{prop-twominima}, to our knowledge,
has never been discussed in literature.\\

It is very possible that a shorter and more direct proof of Proposition~\ref{prop-twominima} exists based on
geometric properties of the family of nested 3-quadrics.

\subsection{The valley of the Higgs potential}

Consider again ${\cal M}^-$, that is the region in the Minkowski space where $\Lambda_{\mu\nu} (r^\mu-m^\mu) (r^\nu-m^\nu) <0$.
For simplicity, we consider here non-singular $\Lambda_{\mu\nu}$.
Let us introduce the {\em valley} $\mathcal{V}$ of the Higgs potential as the intersection of ${\cal M}^-$
with the interior and the surface of $LC^+$. The intersection of ${\cal M}^-$
with the {\em surface} of $LC^+$ will be called the {\em bottom of the valley}.

By construction, $\mathcal{V}$ is the set of all physically realizable points $r^\mu$ that lie strictly deeper than $V_0$.
It follows immediately that if the valley exists, then all local minima of the potential lie in the valley.

The concept of the valley is most useful if the base point $r^\mu=m^\mu$ lies {\em inside}
the future lightcone $LC^+$. In this case, there is the following (almost tauthological) criterion
for the existence of the valley: {\em it exists if and only if $\Lambda_{\mu\nu}$ is not positive definite
in the entire Minkowski space $M$.} Indeed, since $m^\mu$ lies inside $LC^+$, then all points
$r^\mu$ sufficiently close to it (in the sense that all components of $r^\mu-m^\mu$ can be made arbitrarily small)
are physically realizable. Then, by looking at (\ref{Vmunu2}) one sees that in order for the valley to exist,
it is necessary and sufficient that at least one of $\Lambda_i$ is positive.

If the base point $r^\mu=m^\mu$ lies {\em outside} the future lightcone $LC^+$, then the positiveness
of at least one $\Lambda_i$ is necessary, but not sufficient for existence of the valley.
Geometry of ${\cal M}^-$ can be such that it ``misses" the lightcone $LC^+$, so no valley exists.
This, however, can happen only when there is still at least one {\em negative} $\Lambda_i$.
If not, i.e. if all $\Lambda_i>0$, then the valley always exists provided that the EW symmetry is broken,
which can be understood from the above geometric constructions.

The notion of the valley allows one to give a very short proof of non-coexistence of
neutral and charge-breaking minima in any 2HDM (Proposition~3 in \cite{mink}).
Indeed, if $m^\mu$ lies outside $LC^+$, then there is no charge-breaking extremum at all.
If $m^\mu$ lies inside $LC^+$, then consider the valley of the Higgs potential.
If it is absent, then there are no neutral minima, so that the minimum is charge-breaking.
If it is present, then the charge-breaking extremum is a saddle point,
while the minima of the potential must lie on the surface of $LC^+$, corresponding to the
neutral vacuum. Indeed, if some $r^\mu$ lies in the valley, then one can go along the ray from $m^\mu$
passing through $r^\mu$ and still further into the valley, down to its bottom.

\begin{figure}[!htb]
   \centering
\includegraphics[width=5cm]{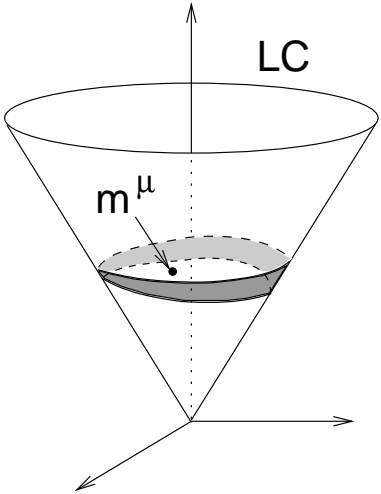}
\caption{$2+1$-dimensional illustration of the non-trivial topology of the valley when $m^\mu$
lies inside $LC^+$. The shaded band is the bottom of the valley on the surface of $LC^+$.}
   \label{fig-valley}
\end{figure}

If the point $r^\mu=m^\mu$ lies inside the future lightcone $LC^+$, then there is room
for {\em non-trivial topology} of the valley. Indeed, since the base point itself,
$r^\mu = m^\mu$, is excluded from $\mathcal{V}$, the topology of $\mathcal{V}$
coincides with the topology of the bottom of the valley. It depends
on the number of positive $\Lambda_i$ and can be understood from the geometric descriptions
given above. In particular,
\begin{itemize}
\item
If all three $\Lambda_i$ are positive, then (the bottom of) the valley is simply connected
and is homotopic to the 2-sphere $S^2$. It has non-trivial second homotopy group
$\pi_2(\mathcal{V}) = \mathbb{Z}$.
\item
If only two of $\Lambda_i$ are positive, the valley is connected, but not simply connected.
It is homotopic to the circle $S^1$ and has
non-trivial fundamental group $\pi_1(\mathcal{V}) = \mathbb{Z}$.
\item
If only one among $\Lambda_i$ is positive,
then the valley is disconnected. Each of its two connected components
is simply connected.
\end{itemize}
In Fig.~\ref{fig-valley} we provide an illustration of the valley in the $2+1$-dimensional case,
which should help visualize the construction.

If the point $r^\mu=m^\mu$ lies {\em outside} the future lightcone $LC^+$, then the
valley has trivial topology. The lightcone $LC^+$ cuts a single line segment
from each ray $\alpha p^\mu$ that belongs to ${\cal M}^-$. Thus, the valley consists of simply connected
regions in $M$.

\subsection{Non-contractible paths in the orbit space}

The non-trivial topology of valley $\mathcal{V}$ of the Higgs potential
allows one to construct non-contractible loops or spheres
in the Higgs orbit space, which follow the bottom of the valley.
They might lead to the existence of metastable quasi-topological configurations
of the vacuum $\langle\phi_i\rangle = \langle\phi_i\rangle(x^\mu)$.
Note that they are possible only in the case when $m^\mu$ lies inside $LC^+$.

If two among $\Lambda_i$ are positive, then the valley is homotopically
equivalent to the circle. This non-contractible loop can gives rise
to the domain wall metastable against spontaneous decay.
In such a wall, in the usual coordinate space, the v.e.v. of the Higgs fields
depend on one of the coordinates, say $x$, $v_1 = v_1(x)$ and $v_2 = v_2(x)$, so that
at $x=\pm \infty$ the v.e.v.'s $v_i$ approach asymptotically their global minimum values, while
in between they follow the corresponding values along the loop.

While going along the loop, one can pass either saddle points or other local minima.
In the simplest case, one passes only one saddle point.
In this case the domain wall separates two regions of the {\em true} vacuum.
If there is another local minimum in the valley, then the domain wall can have
a layered structure with the false vacuum wall sandwiched between
the two high surface tension walls. An interesting case takes place when the two
minima are degenerate, which can happen in the case of
spontaneously violation of a discrete symmetry.

These domain walls were stable enough, they would have intriguing phenomenological properties
via interaction with the fermions.
First, a stable fermions approaching the wall can be trapped inside or reflected back,
since the domain wall will act as an effective thin potential barrier.
The fermions trapped {\em between} the two colliding domain walls can bounce back and forth
accelerating until they can leak outside (i.e. until their wavelength
becomes smaller then the width of the domain wall).

If all three $\Lambda_i$ are positive, then the valley is homotopically equivalent to $S^2$,
and one can have a non-contractible sphere in the Higgs orbit space.
The relevant topological defect would be the string; however, it must be stabilized against
shrinking.

Such topological defects in the two-Higgs-doublet model have been studied
in the literature, see \cite{tomaras}.
Here we would like just to make three comments.
First, the examples discussed there
resulted from the straightforward search in the space of Higgs potential parameters.
In our discussion we gave absolutely general and concise criteria for existence
of such defects in terms of positivity of $\Lambda_i$ and location of $m^\mu$.
Second, our defects involve only scalar fields and not the gauge bosons, and third,
they correspond to non-contractible paths in the space of {\em gauge orbits} rather than
in the space of scalar fields.
It would be interesting to see a quantitative characterization of these configurations
in the reparametrization covariant way.

\section{Discrete symmetries and their spontaneous violation}\label{section-symmetries}

The large number of free parameters in the Higgs potential (\ref{potential}) makes it possible
to introduce into 2HDM new symmetries in addition to the electroweak symmetry.
These are {\em reparametrization symmetries}: they involve not the electroweak
transformations inside the doublets, but transform or mix the doublets themselves.

Investigation of these symmetries, possibility of their spontaneous violation,
as well as their phenomenological consequences is one of the most interesting
aspects of the 2HDM research, see e.g.~\cite{ginzreview}.
The most studied case of such a symmetry is the spontaneous $CP$-violation in 2HDM,
\cite{2HDM,ginzreview,haber2,nishi,nachtmann2}.

Until recently, the study of the presence (or absence) of the spontaneous violation of a discrete symmetry
in 2HDM has been sporadic and was limited to some simple specific cases.
Geometric approach makes it clear that the $CP$-symmetry is just a one specific representative
of a general class of symmetries of the Higgs lagrangian with purely geometric origin,
\cite{mink,group}. A detailed treatment of this more general class of symmetries
was performed in \cite{nachtmann2}.

Here we present an even more general point of view on the
reparametrization symmetries in 2HDM.

\subsection{Classification of explicit reparametrization symmetries possible in 2HDM}

As we have explained in Section~\ref{section-mink}, the Higgs lagrangian
remains invariant under an appropriate simultaneous transformation of fields
{\em and} parameters of the lagrangian. It can happen, however, that
the lagrangian is invariant under some specific transformation
of fields (or parameters) {\em alone}. We call this symmetry
the {\em explicit (reparametrization) symmetry} of the Higgs lagrangian.

In the orbit space, this symmetry corresponds to such a map of the Minkowski space
${\cal M}$ that leaves invariant, separately,
$\Lambda_{\mu\nu} r^\mu r^\nu$, $M_\mu r^\mu$ and kinetic term $K_\mu \rho^\mu$.
The notion of explicit symmetry is invariant
under the Lorentz group of the orbit space transformations.

We start with classification of explicit reparametrization symmetries
that a Higgs lagrangian can possess.
\begin{proposition}\label{prop-classification-symmetries}
Suppose that the Higgs lagrangian is explicitly invariant under some
transformations of $r^\mu$. Let $G$ be the maximal group of such transformations.
Then:\\
(a) $G$ is non-trivial if and only if there exists an eigenvector
of $\Lambda_{\mu\nu}$ orthogonal both to $M_\mu$ and $K_\mu$;\\
(b) group $G$ is one of the following groups: $Z_2$, $(Z_2)^2$,
$(Z_2)^3$, $O(2)$, $O(2)\times Z_2$, or $O(3)$.
\end{proposition}
\begin{proof}
Consider parameters of different parts of the Higgs lagrangian in the prototypical model
(see Section~\ref{section-prototypical}): $\Lambda_{\mu\nu}$,
which is already diagonal, and two four-vectors $M_\mu$ and $K_\mu$.
Let us call their spacelike parts $\Lambda_{ij}$, $M_i$ and $K_i$, respectively.

Any allowed map of ${\cal M}$ that realizes an explicit reparametrization symmetry must preserve
the orbit space $LC^+$.
Let us denote the group of all allowed symmetries of $\Lambda_{\mu\nu}$ by $G_\Lambda$,
and, the groups of all allowed symmetries of $M_\mu$ and $K_\mu$ by $G_M$ and $G_K$, respectively.
Obviously,
\be
G = G_\Lambda \cap G_M \cap G_K\,.
\ee
The allowed symmetry cannot flip the ``timelike'' axis; therefore,
in the frame where $\Lambda_{\mu\nu}$ is diagonal the groups $G_\Lambda$, $G_M$, $G_K$
are in fact the symmetry groups of $\Lambda_{ij}$, $M_i$ and $K_i$, respectively.

Consider now $G_\Lambda$.
If all eigenvalues of $\Lambda_{ij}$ are different, then its only symmetries are
reflections of each of the spacelike eigenaxes.
Such reflections form the group $G_\Lambda=(Z_2)^3$.
If two eigenvalues coincide, then
$G_\Lambda$ is promoted to $O(2)\times Z_2$, and if all three of them are
equal, then $G_\Lambda=O(3)$.
Note that zeros among the eigenvalues of $\Lambda_{\mu\nu}$ do not lead to any additional
reparametrization symmetry.

Note that in all of these cases the following statement holds: if some $Z_2$ group
is a subgroup of $G_\Lambda$, then the generator of this $Z_2$ group
flips the direction of an eigenvector of $\Lambda_{ij}$.

Similarly, $G_M$ is $O(2)$ (rotations around the axis defined by $M_i$), if $M_i$ is a non-zero vector, and $O(3)$ otherwise.
The same holds also for $K_i$, the only difference being the direction of the axis.
It is plain to see that
$$
G_M\cap G_K =
\left\{
\begin{array}{cl}
O(3) & \mbox{if $M_i = K_i = 0$,}\\
O(2) & \mbox{if $M_i$ and $K_i$ are collinear,}\\
Z_2  & \mbox{if $M_i$ and $K_i$ are non-collinear.}
\end{array}
\right.
$$
If we want $G$ to be non-trivial, then the lowest possible symmetry of $M_i$ and $K_i$, $Z_2$,
must be also the symmetry of $\Lambda_{ij}$. With the above remark, it means that
this symmetry flips one of the eigenvectors of $\Lambda_{ij}$. In other words,
both $M_i$ and $K_i$ are orthogonal to this eigenvector. Finally, since this eigenvector
is also the purely spacelike eigenvector of $\Lambda_{\mu\nu}$, we arrive at the first statement
of this Proposition.

Detailed classification depends on the {\em number} of eigenvectors of $\Lambda_{ij}$
that are orthogonal to $M_i$ and $K_i$.
\begin{itemize}
\item
If $M_i$ and $K_i$ are orthogonal to all three eigenvectors, which can be possible only
when they both are zero vectors, then $G=G_\Lambda$.
\item
If $M_i$ and $K_i$ are orthogonal to two eigenvectors, which is possible
only when $M_i$ and $K_i$ are collinear and are themselves eigenvectors of $\Lambda_{ij}$, then
$G = (Z_2)^2$ or $O(2)$.
\item
Finally, if there is only one eigenvector of $\Lambda_{ij}$ orthogonal both to $M_i$ and $K_i$,
then the symmetry group is $Z_2$.
\end{itemize}
\end{proof}

The necessary and sufficient condition formulated in Proposition~\ref{prop-classification-symmetries}a
can be written in a reparametrization-invariant way. The method is essentially the same
as in \cite{group}. We introduce
\be
K_{0\mu} \equiv K_\mu\,,\quad
K_{1\mu} \equiv \Lambda_{\mu}{}^\nu K_\nu\,,\quad
K_{2\mu} \equiv (\Lambda^2)_{\mu}{}^\nu K_\nu\,,\quad
K_{3\mu} \equiv (\Lambda^3)_{\mu}{}^\nu K_\nu\,,
\ee
where $\Lambda^k$ is the $k$-th power of $\Lambda_{\mu\nu}$.
The same series can be written for $M_\mu$.
For any four four-vectors $a^\mu$, $b^\mu$, $c^\mu$, and $d^\mu$ we introduce
the short-hand notation
$$
(a,b,c,d) \equiv \epsilon_{\mu\nu\rho\sigma} a^{\mu} b^{\nu} c^{\rho} d^{\sigma}\,.
$$
Then the condition ``there exists an eigenvector of $\Lambda_{\mu\nu}$ orthogonal to $K_\mu$''
can be written as
\be
(K_{0}, K_{1}, K_{2}, K_{3}) = 0\,.\label{epsilon4k}
\ee
Note that since $K^\mu$ always lies inside the future lightcone, it can be orthogonal
only to spacelike eigenvectors of $\Lambda_{\mu\nu}$, which is exactly what is needed.
Then, the statement of Proposition~\ref{prop-classification-symmetries}a can be reproduced if
we accompany (\ref{epsilon4k}) with the similar condition for $M_\mu$,
\be
(M_{0}, M_{1}, M_{2}, M_{3}) = 0\,,\label{epsilon4m}
\ee
and the condition that these two 4-vectors be orthogonal to the {\em same} eigenvector of $\Lambda_{\mu\nu}$,
for example:
\be
(M_{0}, M_{1}, M_{2}, K_{0}) = 0\,,\label{epsilon3m1k}
\ee
Note that these conditions can be straightforwardly checked in any frame of reference
once $\Lambda_{\mu\nu}$, $M_\mu$, and $K_\mu$ are known,
although their relation with the conditions formulated in
\cite{haber2,group,nishi} might be complicated.

\subsection{Phenomenologically interesting discrete symmetries}

Let us focus now on the situation when all $\Lambda_i$ are distinct,
which means that there can be only discrete explicit symmetries.
According to the above Proposition, this group can be $(Z_2)^k$ with $k=1,2,3$, and
in the diagonal basis is generated by flipping of the eigenaxes of $\Lambda_{ij}$.

In the representation (\ref{ri}), flips of the axes
correspond to the following transformations of the Higgs fields (index $\alpha$ indicates
the upper and lower components in each the doublet):
\bea
\mbox{flip of the first axis:} && \phi_{1\alpha}\to\phi_{1\alpha}^*\,,\quad \phi_{2\alpha}\to -\phi_{2\alpha}^*\,,\nonumber\\
\mbox{flip of the second axis:}&& \phi_{1\alpha}\to\phi_{1\alpha}^*\,,\quad \phi_{2\alpha}\to \phi_{2\alpha}^*\,,\nonumber\\
\mbox{flip of the third axis:} && \phi_{1\alpha}\leftrightarrow\phi_{2\alpha}^*\,.
\eea
The discrete symmetries that are usually discussed in the context of 2HDM can be constructed from these
elementary blocks.
\begin{itemize}
\item
{\bf Explicit $CP$-conservation} takes place when, after an appropriate reparametrization,
all parameters of the Higgs lagrangian are real. It means that in this basis Im$(\phi_1^\dagger \phi_2)$
does not appear in the lagrangian. This situation corresponds
precisely to the Higgs lagrangian being symmetric under the flipping the second axis.
In the diagonal basis, this symmetry takes place
if $M_\mu$ and $K_\mu$ have their second components equal to zero.
\item
What is conventionally called the {\bf explicit $Z_2$-symmetry} of the Higgs potential
is the invariance under transformation $\phi_1\to\phi_1\,,\ \phi_2\to -\phi_2$
(which implies $m_{12}^2=0$ and $\lambda_6 = \lambda_7 = 0$).
It corresponds to the simultaneous flipping of first and second axes.
The only way to have this symmetry in 2HDM is to require that vectors $M_i$ and $K_i$
be invariant under {\em separate} flips of first and second axes.
In other words, it corresponds to the Higgs potential with the symmetry group
at least $Z_2\times Z_2$ (so, the standard terminology here is a misnomer).
In the diagonal basis, it implies that $M_i$ and $K_i$
are both aligned along the third axis.
\end{itemize}

One can say that explicit $CP$-conservation serves as a ``prototypical'' case
of the explicit $Z_2$ symmetry, while what is conventionally called ``$Z_2$-symmetry''
serves as a ``prototypical'' case of the $Z_2\times Z_2$ symmetry.
It means also that the conventional ``$Z_2$-symmetry'' immediately implies
explicit $CP$-conservation.

\subsection{Symmetries of potential vs. symmetries of lagrangian}\label{section-symm-pot-lang}

It has been noted above that the exact value of the kinetic four-vector $K_\mu$
is not important when one studies the general structure of the extrema of the 2HDM orbit space.
Thus, it makes sense to distinguish the symmetries of the Higgs {\em lagrangian},
which is what we just discussed, and the symmetries of the Higgs {\em potential}.
The latter is given by the group $G_\Lambda \cap G_M$ and can be larger than $G$.

A very simple case of a potential whose symmetry group is larger than the symmetry
group of the entire Higgs lagrangian is
\be
V = 16\lambda \left[(\fd_1\f_1)-{v^2\over 2}\right]^2 + \lambda \left[(\fd_2\f_2)-{4v^2\over 2}\right]^2\,.
\ee
This potential is symmetric under $\phi_2 \leftrightarrow 2\phi_1$, while
the kinetic term of the Higgs lagrangian is not.

We stress that the explicit symmetries of the potential are more important
for the study of the general structure of the vacuum in 2HDM than the symmetries
of the entire Higgs lagrangian. Section~\ref{section-global} provides
an illustration of its importance.

\subsection{What is the maximal spontaneous violation of a discrete symmetry?}

Even if the Higgs potential is invariant under some transformation of $\Phi$,
the vacuum expectation values $\langle \Phi\rangle$ do not necessarily
have to respect the same symmetry. In the orbit space of 2HDM, if the Higgs potential
is invariant under group $G$ of transformation of $r^\mu$,
then the position of the global minimum,
$\langle r^\mu\rangle$, might be invariant only under the proper subgroup of $G$.
In such situations one talks about spontaneous violation of the symmetry.
Note that the {\em set} of all minima still respects the explicit
reparametrization symmetry.

Again, let us focus on the generic situation, i.e. when all eigenvalues of $\Lambda_{\mu\nu}$
are non-zero and different. The symmetry group of the potential is then $(Z_2)^k$ with $k=1,2$, or 3.
An interesting question is: what is the maximal violation of the explicit symmetry in 2HDM?
The answer is given by the following Proposition:
\begin{proposition}\label{prop-spontaneous-violation}
The maximal spontaneous violation of an explicit discrete symmetry of the 2HDM potential
or lagrangian consists in removing one $Z_2$ factor.
\end{proposition}
\begin{proof}
Let us start with the spontaneous violation of a discrete symmetry of the Higgs {\em potential}.

A 2HDM Higgs potential with an explicit $(Z_2)^k$ symmetry, with $k=1,\,2$, or 3,
implies that in the diagonal basis there are exactly $k$ eigenaxes along which
$M_i$ has zero components. The question is how many zero components $\lr{r^\mu}$ can have in this basis.

First, note that the charge-breaking extremum never breaks the explicit symmetry.
It follows from the fact that there can be only one charge-breaking minimum and the above remark
that the set of all minima is invariant under the explicit
symmetry transformations.

Turning to the neutral vacuum, recall the equation for a neutral extremum of the potential
\be
\Lambda_{\mu\nu} \lr{r^\nu} - \zeta \cdot \lr{r_\mu} = M_\mu\,,\label{solution}
\ee
with some real parameter $\zeta$.
Vector $\lr{r^\mu}$ is restricted to lie on the surface of the forward lightcone $LC^+$.
The crucial fact is that the surface of $LC^+$ is a manifold with codimension 1.
It means that when we search for an extremum located on the surface of $LC^+$,
we need to introduce {\em only one} Lagrange multiplier $\zeta$ in (\ref{solution}).

Now, let us rewrite (\ref{solution}) in the diagonal basis:
$$
(\Lambda_0-\zeta)\cdot \lr{r_0}=M_0\,,\quad (\Lambda_i-\zeta)\cdot \lr{r_i}=M_i\,.
$$
Recall that $k$ components of $M_i$ are zeros.
The least possible numbers of zeros in among the coordinates of $\lr{r_i}$ is $k-1$.
Indeed, one can adjust $\zeta$ equal to one of $\Lambda_i$ so that
the corresponding component of $\lr{r_i}$ can be non-zero.
Since all $\Lambda_i$ are different, then all other $\Lambda_i-\zeta$
are non-zero, and the corresponding components of $\lr{r_i}$ must be set to zero.
Thus, the symmetry of $\lr{r^\mu}$ is lower than the symmetry of the potential by
a single $Z_2$ factor.

Turning now to the spontaneous violation of a discrete symmetry of the Higgs {\em lagrangian}, note
that in this case the symmetry group of the potential alone is $(Z_2)^n$, where $n\ge k$,
while $(Z_2)^k$ is the common symmetry of the potential and the kinetic term.
The symmetry of the potential can be broken spontaneously down to $(Z_2)^{n-1}$, so the symmetry
of the $\lr{r^\mu}$ is at least $(Z_2)^{k-1}$.
\end{proof}

This Proposition has immediate consequences for establishing
the conditions of spontaneous $CP$-violation.
The vacuum of 2HDM can spontaneously violate $CP$-symmetry,
if and only if there are {\em no} discrete symmetries under which $\langle r^\mu\rangle$
is invariant. Indeed, if there were even a single $Z_2$ factor, then by redefinition
of the Higgs fields one would arrive at $\langle r^\mu\rangle$ in the form of
$(\cdot,\,\cdot,\,0,\,\cdot)$, where $\cdot$ labels a generic value.
This means that it would be possible to perform a reparametrization transformation that removes
the relative phase between the v.e.v.'s of the doublets.

But according to this Proposition, this can take place only when the
group of the explicit symmetries of the Higgs lagrangian in {\em exactly} $Z_2$.
Too symmetric Higgs lagrangian, with $G$ larger than $Z_2$, cannot lead to spontaneous
$CP$-violation. This particular conclusion was also reached in \cite{mink}.

Note that Proposition~\ref{prop-spontaneous-violation} related the ``strength''
of spontaneous violation of discrete symmetries to the geometry of the {\em strata} of the
2HDM orbit space. Roughly speaking, a stratum can be defined as a set of points of the orbit
space that can be connected by an extended reparametrization transformation.
In 2HDM, the groups of extended reparametrization transformation is $GL(2,C)$,
which induces proper Lorentz group times dilatations in the orbit space.
We thus obtain three strata: the vertex of the cone $LC^+$, the surface of $LC^+$
and the interior of $LC^+$, \cite{sartori2hdm}.

It is the surface of $LC^+$ (manifold with codimension 1) that happens to correspond to
neutral vacua. If there were other strata with codimension $p$, then
one would need $p$ Lagrange multipliers, and then the spontaneous violation
could reduce the explicit symmetry by $(Z_2)^p$.

Finally, the fact that the surface of $LC^+$ is a manifold with codimension 1, is related to the
very nature of the electroweak symmetry breaking. In EWSB we reduce the initial four-dimensional
$SU(2)\times U(1)$ electroweak symmetry to the one-dimensional $U(1)_{EM}$ symmetry.
The codimension 1 of the boundary of $LC^+$ comes from the one degree of freedom of the
remaining symmetry.

It appears that the relation between the maximal strength of the spontaneous violation of discrete symmetries
and the dimension of the remaining symmetry after EWSB
is not specific to 2HDM but is more universal.

\section{The global minimum and its bifurcation}\label{section-global}

In our discussion of the global minimum of the potential we assume as usual that $\Lambda_{\mu\nu}$
has already been diagonalized.
As it was shown in \cite{mink,nachtmann}, there can be up to six neutral extrema
of the Higgs potential of a generic 2HDM. Let us fix $\Lambda_{\mu\nu}$ and
change $M_\mu$. As parameters of the potential change, the positions
and depths of these extrema will continuously change until
a bifurcation occurs, when several extrema merge or or one extremum splits.

In our previous analysis in \cite{mink} we did not distinguish the global minimum
from the other extrema, and no method was proposed of how to recognize when
it is the global minimum that bifurcates, and not the other extrema.
Here, we fill this gap with the aid of the above geometric constructions.\\

The starting point is the fact that the depth of the global minimum
is given by such a 3-surface ${\cal M}^{C_\mathrm{min}}$ that barely touches, but never intersects,
the future lightcone $LC^+$. This makes the global minimum distinct from the other extrema,
whose ${\cal M}^C$ not only touch, but also intersect $LC^+$.

Let us study the properties of the contact between ${\cal M}^{C_\mathrm{min}}$ and $LC^+$.

We first note that ${\cal M}^{C_\mathrm{min}}$ and $LC^+$
can touch in {\em not more than two points}. Indeed, each of these
3-surfaces is a quadric. Intersection of two 3-quadrics is described by fourth degree polynomials.
Each contact point is a degenerate case of a sphere with zero radius and requires
at least a two-degree polynomial. Thus, a fourth degree polynomial can define
no more than two contact points.

Alternatively, one could simply apply our Proposition~\ref{prop-twominima}.

The immediate consequence is that the 2HDM vacuum cannot be
degenerate more than twice. The question now arises: {\em when} can it be degenerate?
The answer is given by the following Proposition:

\begin{proposition}\label{prop-degenerate-vacuum}
The vacuum can be twice degenerate only as a result of spontaneous violation of a
discrete $Z_2$ symmetry of the potential.
\end{proposition}
\begin{proof}
Let us first introduce a definition.
Let $P$ be a quadric in the (pseudo)Euclidean space $\mathbb{R}^n$ defined by equation
\be
P(x_i)=a_{ij}x_ix_j + 2b_i x_i + c = 0\,,\quad x_i \in \mathbb{R}^n\,.
\ee
We call two quadrics $P$ and $P'$ {\em aligned}, if the corresponding matrices $a_{ij}$
and $a^\prime_{ij}$ have the same eigenvectors. In plain words,
quadrics $P$ and $P'$ are oriented in the same directions, although they can be shifted
in respect to each other.

In the frame where $\Lambda_{\mu\nu}$ is diagonal,
the 3-quadrics ${\cal M}^{C_\mathrm{min}}$, whose equation is
\be
\Lambda_0 p_0^2 - \Lambda_1 p_1^2 - \Lambda_2 p_2^2 - \Lambda_3 p_3^2 = C_\mathrm{min}\,,
\ee
and the forward lightcone $LC^+$ are aligned.

In Appendix~\ref{appendixquadric} we prove Lemma~\ref{lemma-quadrics}, which states that if two aligned quadrics
have exactly two contact points, then they have a common $Z_2$ symmetry, which consists
in reflection of one of the axes. The two points are mapped onto each other by this
reflection; so, they have all the coordinates equal except the one that transforms
under the reflection.

The properties of ${\cal M}^{C_\mathrm{min}}$ are defined by the parameters of the potential:
its shape is given by eigenvalues $\Lambda_0,\,\Lambda_i$, while the position of its symmetry
center is given by $m^\mu$. The statement of Lemma~\ref{lemma-quadrics} implies,
in our language, that $m^\mu$ (and, therefore, $M^\mu$) lies in the 3-plane orthogonal
to one of the eigenvectors of $\Lambda_{\mu\nu}$. That is, an explicit discrete symmetry
of the Higgs potential is realized in this coordinate frame.

The fact that the contact points (i.e. the values of $\lr{r^\mu}$ that realize the global minimum)
{\em do not} lie in the above mentioned 3-plane means that the vacuum does not possess
this symmetry. In other words, this symmetry is spontaneously violated.
\end{proof}

Note that this Proposition deals with the symmetries of the potential, not of the
entire Higgs lagrangian. It might happen that the Higgs lagrangian does not have any
discrete symmetry at all and still has the twice degenerate minimum. The Proposition just proved
affirms that in this case the {\em potential} has a certain hidden symmetry, which might be not obvious
from the simple inspection of the lagrangian.

Clearly, the Proposition just proved also implies that the Higgs potential with two
{\em nearly} degenerate minima necessarily implies existence of an {\em approximate} symmetry of the potential.

As shown in \cite{mink}, multiple minima can take place only when $m_\mu$ lies inside certain caustic cones.
If one starts with the double-minimum configuration, fixes $m_0$ and increases spacelike
coordinates $m_i$, $m_i \to \alpha m_i$,
then the two minima approach each other and at some point the double minimum plus a saddle point
merge into a single minimum. Geometrically, this is the four-point contact of $LC^+$ and ${\cal M}^{C_\mathrm{min}}$.
Passing though such a point leads to the bifurcation of the extrema of the
potential\footnote{Here we only mean that a bifurcation happens upon continuous change of the
free parameters of the potential. Whether it corresponds to a real finite-temperature phase transition
requires further study.}.

In principle, not only denegerate global minima but also degenerate saddle points
lead to an explicit discrete symmetry. The following Proposition gives the criterion,
when it is the global minimum, not just an arbitrary stationary point,
than experiences the bifurcation:
\begin{proposition}\label{prop-degenerate-vacuum2}
The global minimum exhibits spontaneous violation of the $Z_2$ symmetry along
the $k$-th eigenaxis of $\Lambda_{\mu\nu}$ only if the corresponding eigenvalue $\Lambda_k$
is positive and is the largest spacelike eigenvalue.
\end{proposition}
\begin{proof}
Let us first show that if all $\Lambda_i$ are negative, then there can be no bifurcation,
and hence the vacuum cannot be degenerate.

First, it is obvious that two convex non-intersecting bodies
cannot touch in two and only two separate points. Indeed, if they touch in two points,
then all points lying on the line segment between them belong to each
of the two convex bodies. So, either they touch along a line segment or they intersect.

The forward lightcone $LC^+$ together with its interior is a convex body.
If all $\Lambda_i <0$ then ${\cal M}^{<C_\mathrm{min}}$ is a 3-ellipsoid,
which is also a convex body. When looking for the neutral vacua, we are interested in the situation
when $m^\mu$ lies outside the $LC^+$ (otherwise,
the global minimum would be the charge-breaking one).
Thus, $LC^+$ and ${\cal M}^{C_\mathrm{min}}$
touch but do not intersect. Since both are convex and since ${\cal M}^{C_\mathrm{min}}$
does not contain any line segment, they cannot touch in more than a single point,
so there can be no bifurcation of the global minimum in this case.

\begin{figure}[!htb]
   \centering
\includegraphics[width=12cm]{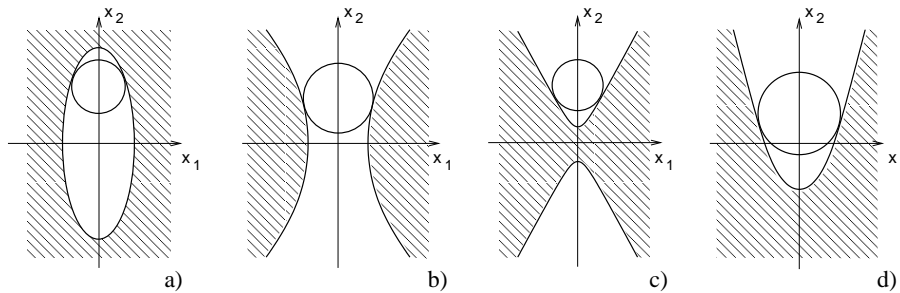}
\caption{The four possible cases of how the two-dimensional section (spanned by $e_1$ and $e_2$)
of ${\cal M}^{C_\mathrm{min}}$ and $LC^+$ can look like;
(a): $\Lambda_1,\, \Lambda_2>0$;
(b) and (c): $\Lambda_1>0$, $\Lambda_2<0$;
(d): $\Lambda_1 >0,\, \Lambda_2=0$.
In all cases $\Lambda_1>\Lambda_2$. The shaded region is the section of ${\cal M}^{<C_\mathrm{min}}$.
}
   \label{fig-osculating}
\end{figure}

Now suppose that at least one $\Lambda_i$ is positive. Then, according to discussion in
Section~\ref{sectequipotential}, any 3-surface ${\cal M}^{C}$ is a 3-hyperboloid
or 3-paraboloid, whose shape and topology
depend on the signs of $\Lambda_i$ and on $C$.

Consider now the two contact points of ${\cal M}^{C_\mathrm{min}}$ and $LC^+$.
According to Lemma~\ref{lemma-quadrics}, they lie symmetrically in respect to the common
symmetry 3-plane of ${\cal M}^{C_\mathrm{min}}$ and $LC^+$. Let us denote the coordinate orthogonal
to this plane as $x_1$. Then, the two contact points are
$$
r^\mu_+ = (x_0,\,x_1,\,x_2,\,x_3)\quad \mbox{and}\quad r^\mu_- = (x_0,\,-x_1,\,x_2,\,x_3)\,.
$$
Consider now a two-dimensional section of the two quadrics ${\cal M}^{C_\mathrm{min}}$ and $LC^+$
by a plane than passes through these two contact points and spanned by the eigenvectors $e_1$ and $e_2$.
The section of $LC^+$ by this plane gives a circle, while the section
of ${\cal M}^{C_\mathrm{min}}$ can yield an ellipse/parabola/hyperbola, which
has two common contact points with the circle.

There are four possibilities to be considered, which are shown in Fig.~\ref{fig-osculating}.
In each case, the shaded region corresponds to the section of ${\cal M}^{<C_\mathrm{min}}$;
by definition, the circle must be disjoint from this region.
By direct inspection one can see that in all three cases $\Lambda_1 > \Lambda_2$.
Now, one can repeat the same check for another section, spanned by eigenvectors $e_1$ and $e_3$,
and obtain $\Lambda_1 > \Lambda_3$.

The overall conclusion is that if the global minimum exhibits spontaneous violation of the
$Z_2$ symmetry generated by the flip of the $k$-th axis, then $\Lambda_k$ must be
positive and be the largest spacelike eigenvalue of $\Lambda_{\mu\nu}$.
\end{proof}

This Proposition shows, in particular, that in spontaneous $CP$-violation
can take place only when $\Lambda_2$ is positive and is larger than $\Lambda_1$, $\Lambda_3$.
This result was also found in \cite{mink} by straightforward algebra.

\section{Conclusions and outlook}\label{section-conclusions}

The aim of this paper is to deepen the geometric understanding of the phenomena
that can happen in a general 2HDM. Following the Minkowski-space approach introduced in \cite{mink},
we investigated the geometric properties of the Higgs potential in the orbit space
and its minima.

We introduced the equipotential surfaces in the orbit space and showed that they are intersections
of two 3-quadrics in the Minkowski space ${\cal M}$.
The search for the global minimum was reformulated as the search of such a 3-quadric
that touches but never intersects the forward lightcone $LC^+$.

This reformulation led us to several observations about the minima of any 2HDM.
Namely, we proved that if 2HDM has a discrete set of minima, then it cannot have more than
two minima. This means, in particular, than the 2HDM with the explicit $CP$-symmetry
cannot have simultaneously $CP$-conserving and $CP$-violating minima.
These statements are in contradiction with the results of numerical studies
reported in \cite{twominima}.
We also proved that if the global minimum happens to be doubly degenerate,
then it can take place only as a result of spontaneous breaking
of a certain $Z_2$ symmetry of the potential. The eigenvalue of $\Lambda_{\mu\nu}$
associated with this symmetry must be the largest among all the spacelike eigenvalues.

We defined the valley of the Higgs potential and discussed its topological properties.
In particular, we observed that non-trivial topology of the valley makes it possible
to construct non-contractible loops in the Higgs orbit space, leading
to metastable topological configurations (either walls or strings)
purely within the scalar sector of 2HDM. We gave concise reparametrization-invariant
criteria when such configurations can take place.

We also discussed discrete symmetries of 2HDM from a more general
point of view than is usually done. We discussed differences between explicit symmetries
of the Higgs potential and the entire Higgs lagrangian and
gave their complete classification. We also found what the maximal spontaneous violation
of a discrete explicit symmetry consists in removing only one $Z_2$ factor,
which is related to the residual symmetry after EWSB.\\

The geometric contructions introduced in this work are not specific for the $3+1$-dimensional
geometry. Hopefully, one can apply them to the analysis of the general $N$-Higgs doublet model,
whose analysis was started in \cite{nishi,nishiN,barrosoN}.\\

I am thankful to Ilya Ginzburg and Celso Nishi for discussions and useful comments.
This work was supported by FNRS and partly by grants RFBR 05-02-16211 and NSh-5362.2006.2.

\appendix

\section{Manipulation with 4-tensor $\Lambda_{\mu\nu}$}\label{appendixlambda}

Here we collect some simple facts about the real symmetric 4-tensor $\Lambda_{\mu\nu}$.

Let us first give explicit expressions for $\Lambda_{\mu\nu}$ with raised indices:
\be
\Lambda_{\mu\nu} =
\left(\begin{array}{cc}
\Lambda_{00} & \Lambda_{0j} \\
\Lambda_{0i} & \Lambda_{ij}
\end{array}
\right)\,,\quad
\Lambda_{\mu}{}^{\nu} = \Lambda_{\mu\alpha} g^{\alpha\nu} =
\left(\begin{array}{cc}
\Lambda_{00} & - \Lambda_{0j} \\
\Lambda_{0i} & - \Lambda_{ij}
\end{array}
\right)\,,\quad
\Lambda^{\mu\nu} =
\left(\begin{array}{cc}
\Lambda_{00} & - \Lambda_{0j} \\
- \Lambda_{0i} & \Lambda_{ij}
\end{array}
\right)\,.\label{lambdamunu3}
\ee
Note that $\Lambda_{\mu}{}^\nu$ is not symmetric anymore.

The eigenvalues $\Lambda_i$ and eigenvectors $e_{(i)}^\mu$ of $\Lambda_{\mu\nu}$ are defined according to
\be
\Lambda_{\mu\nu} e_{(i)}^\nu = \Lambda_i\, g_{\mu\nu} e_{(i)}^\nu\,,\qquad
\Lambda_{\mu}{}^{\nu} e_{(i)\, \nu} = \Lambda_i\, e_{(i)\, \mu}\,.\label{eigen}
\ee
Note the presence of $g_{\mu\nu}$ in the first line here. The fact that $\Lambda_{\mu}{}^{\nu}$
is not symmetric means that the (spacelike) eigenvalues will be, in general, complex.
However, as it was proved in \cite{mink}, positive definiteness of $\Lambda_{\mu\nu}$
on and inside the forward lightcone $LC^+$ makes the spacelike eigenvalues real and smaller than $\Lambda_0$.

In the diagonal basis, one has:
$$
\Lambda_{\mu\nu} =
\left(\begin{array}{cccc}
\Lambda_0 & 0 & 0 & 0\\
0 & -\Lambda_1 & 0 & 0\\
0 & 0 & -\Lambda_2 & 0\\
0 & 0 & 0 & -\Lambda_3 \end{array}\right)\,,\quad
\Lambda_{\mu}{}^{\nu} =
\left(\begin{array}{cccc}
\Lambda_0 & 0 & 0 & 0\\
0 & \Lambda_1 & 0 & 0\\
0 & 0 & \Lambda_2 & 0\\
0 & 0 & 0 & \Lambda_3 \end{array}\right)\,.
$$
If one consider a quadratic form in the space of 4-vectors $p^\mu$ constructed on $\Lambda_{\mu\nu}$,
then in the diagonal basis it looks as
$$
\Lambda_{\mu\nu} p^\mu p^\nu = \Lambda_0 p_0^2 - \sum_i \Lambda_i p_i^2\,.
$$
This quadratic form is positive definite in the entire space of non-zero vectors $p^\mu$, if and only if
all $\Lambda_i$ are {\em negative}. One could think of $\Lambda_{\mu\nu}$ as defining a new metric in the
space of vectors $p^\mu$. If all $\Lambda_i$ are negative, this metric has the usual euclidean signature.

\section{Quadrics with two contact points}\label{appendixquadric}

Here we prove the lemma that was used in Propositions~\ref{prop-degenerate-vacuum}
and~\ref{prop-degenerate-vacuum2}.

Let $P$ be a quadric in the Euclidean space $\mathbb{R}^n$ defined by equation
\be
P(x_i)=a_{ij}x_ix_j + 2b_i x_i + c = 0\,,\quad x_i \in \mathbb{R}^n\,.
\ee
We call two quadrics $P$ and $P'$ {\em aligned}, if the corresponding matrices $a_{ij}$
and $a^\prime_{ij}$ have the same eigenvectors, or, in plain words,
if quadrics $P$ and $P'$ are oriented in the same directions (although they can be shifted
in respect to each other).

Two $n-1$-dimensional quadrics can intersect along a fourth-order $n-2$-dimensional manifold in $\mathbb{R}^n$.
In special cases the intersection reduces just to two isolated contact points.
Here, the contact point, in contrast to the intersection point,
means that the two quadrics not only pass through this point, but also have
parallel normals at this point.
\begin{lemma}\label{lemma-quadrics}
If two aligned quadrics $P$ and $P'$ in $\mathbb{R}^n$ have exactly two contact points,
then $P$ and $P'$ have a common $Z_2$ symmetry,
and the two contact points are mapped onto each other under this symmetry.
\end{lemma}
\begin{proof}
The proof will go as follows. We will consider the two contact points
together with the normals at these points as some ``initial data'' and will
proceed by reconstructing the quadrics that satisfy these data.
We will find that if two different quadrics
satisfying these data are aligned,
then they must have a common symmetry and
the initial data are symmetric.

Let us first choose the coordinate frame in which the two contact points are
$$
x^{\pm}_i = (\pm x_1,\, 0,\, \dots,\, 0)\,,
$$
where $x_1 \not = 0$. The generic equation of the quadric is
\be
P(x_i) = a_{ij}x_ix_j + 2b_i x_i + c = 0\,,\label{genericquadric}
\ee
with some symmetric non-zero $a_{ij}$.
Here, $c$ is also non-zero, because the origin of the chosen coordinate
frame does not belong to the quadrics
(otherwise, a line would intersect a quadric at three points).
Thus, we can always set $c=1$ in (\ref{genericquadric}).

The fact that the quadric goes through both $x^+$ and $x^-$ leads to
\be
a_{11}x_1^2 = 1\,,\quad b_1 = 0\,.
\ee
The normals to the quadric at $x^\pm$ are defined by
$$
t_i \equiv \partial P/\partial x_i = 2 a_{ij}x_j + 2b_i\,.
$$
The first coordinate of $t_i$ at the contact points $x^\pm$ is
$t_1^\pm = \pm 2a_{11} x_1 = \pm 2/x_1$, while all the other coordinates
are
$$
t_i^\pm = \pm 2 a_{1i} x_1 + 2b_i\,,\quad i\not = 1\,.
$$
Alternatively, the direction of $t_i^\pm$ can be given by coefficients
$c^{\pm}_{(i)}$ defined via:
\be
\pm 2a_{11} x_1 c^{\pm}_{(i)} = \pm 2 a_{1i} x_1 + 2b_i\,, \quad i \not = 1\,.
\ee
If these coefficients are known, then the parameters of the quadric can be
written as:
\be
b_i = {2 \over x_1} (c^{+}_{(i)} - c^{-}_{(i)})\,,\quad
a_{1i} = {1 \over 2x_1^2} (c^{+}_{(i)} + c^{-}_{(i)})\,, \quad i \not = 1\,.
\ee
In other words, if the initial data ($x_1$ and the values of $c^{\pm}_{(i)}$)
are given, then all $b_i$ and $a_{1i}$ (including $i=1$) are
uniquely reconstructed, while $a_{ij}$ for $i,j \not = 1$ can be chosen at will.

Now, suppose we have two quadrics that satisfy these initial data,
whose $a_{ij}$ and $a'_{ij}$ can differ only for $i,j\not = 1$.
The alignment of the two matrices $a_{ij}$ and $a'_{ij}$ is equivalent to
$[a,a']=0$.

Consider first the case of completely symmetric initial data, which implies
$$
c^{+}_{(i)} + c^{-}_{(i)} = 0 \quad \forall i\not = 1\,.
$$
This leads to $a_{1i} = 0\ \forall i\not = 1$. Together with $b_1 = 0$,
it makes eq.~(\ref{genericquadric}) symmetric under change of the sign
of $x_1$, which generates the required $Z_2$ symmetry.
In other words, {\em all} quadrics that satisfy the same symmetric initial data
are symmetric under the same $Z_2$ symmetry. In particular,
this family of quadrics contains pairs of completely aligned quadrics
with all eigenvalues $\tilde a_{i} \not = \tilde a'_{i}\ \forall i\not = 1$.
The last requirement is essential because if
$\tilde a_{m} = \tilde a'_{m}$ for some $m \not = 1$, then
the two quadrics will touch not in two points, but along a whole second-order curve.

Now, suppose that the initial data are not completely symmetric.
It means that at least for some $m$, $c^{+}_{(m)} + c^{-}_{(m)} \not = 0$.
In this case, explicit calculations show that the commutator $[a,a']$
can be zero only at the expense of having equal eigenvalues
$\tilde a_{m} = \tilde a'_{m}$ for some $m \not = 1$.
Again, this leads to the contact along a whole curve, which contradicts
the assumption.

The conclusion is that the only way for two aligned quadrics to
touch exactly in two points is to do it symmetrically.
\end{proof}

\end{document}